\documentclass[fleqn,10pt]{wlscirep}
\title{Versatility of nodal affiliation to communities}
\usepackage{amssymb}
\usepackage{xspace}
\newcommand{\SU}{\texttt{SU}\xspace}
\newcommand{\EU}{\texttt{EU}\xspace}
\newcommand{\TU}{\texttt{TU}\xspace}
\newcommand{\SC}{\texttt{SC}\xspace}
\newcommand{\EC}{\texttt{EC}\xspace}
\newcommand{\TC}{\texttt{TC}\xspace}
\newcommand{\E}{\operatorname{\mathbb{E}}}
\newcommand{\citep}{\cite}
\newcommand{\citet}{\cite}
\newcommand{\CM}{\ensuremath{\checkmark}}

\author[1,*]{Maxwell Shinn}
\author[1]{Rafael Romero-Garcia}
\author[1,2]{Jakob Seidlitz}
\author[1]{Franti\v{s}ek V\'{a}\v{s}a}
\author[1]{Petra E.~V\'{e}rtes}
\author[1,3,4]{Edward Bullmore}

\affil[1]{Department of Psychiatry, Behavioural and Clinical Neuroscience Institute, University of Cambridge, Cambridge CB2 0SZ, United Kingdom}
\affil[2]{Developmental Neurogenomics Unit, National Institute of Mental Health, Bethesda, MD 20892, USA}
\affil[3]{GlaxoSmithKline Clinical Unit Cambridgeshire \& Peterborough NHS Foundation Trust, Cambridge, Addenbrookes Hospital, Cambridge CB2 0QQ, United Kingdom}
\affil[4]{GlaxoSmithKline R\&D, Immunology \& Inflammation Therapeutic Area Unit, Stevenage SG1 2NY, United Kingdom}
\affil[*]{maxwell.shinn@yale.edu}

\begin{abstract}
  Graph theoretical analysis of the community structure of networks
  attempts to identify the communities (or modules) to which each node
  affiliates.  However, this is in most cases an ill-posed problem, as
  the affiliation of a node to a single community is often ambiguous.
  Previous solutions have attempted to identify all of the communities
  to which each node affiliates.  Instead of taking this approach, we
  introduce \emph{versatility}, $V$, as a novel metric of nodal
  affiliation: $V \sim 0$ means that a node is consistently assigned
  to a specific community; $V \gg 0$ means it is inconsistently
  assigned to different communities.  Versatility works in conjunction
  with existing community detection algorithms, and it satisfies many
  theoretically desirable properties in idealised networks designed to
  maximise ambiguity of modular decomposition.  The local minima of
  global mean versatility identified the resolution parameters of a
  hierarchical community detection algorithm that least ambiguously
  decomposed the community structure of a social (karate club) network
  and the mouse brain connectome.  Our results suggest that nodal
  versatility is useful in quantifying the inherent ambiguity of
  modular decomposition.
\end{abstract}
\begin{document}

\flushbottom
\maketitle

\thispagestyle{empty}

\section{Introduction}

The community structure of a network divides the network into groups,
or communities, which share topological similarity.  These communities
are most commonly defined to be non-overlapping groups which maximise
the strength of edges within the community and minimise the strength
of edges which leave the community, where each node is a member of one
and only one community.

Sometimes, the community structure of a network is evident even to
an untrained observer.  It is very clear which nodes belong to which
community, and which nodes and edges are involved in linking
communities together.  In other words, the overall community structure
is unambiguous for nearly all of the nodes in the network.

However, in most networks, the modular decomposition of community
structure is an ill-posed problem, as not all nodes can be assigned
unambiguously to a single community. Techniques previously developed
to deal with this situation include algorithms that allow overlapping
communities \citep{xie2013} and algorithms that work not with
communities themselves, but rather with probability distributions of
communities via multi-layer networks \citep{betzel2015}. These
approaches, while attractive in theory, can be challenging to
operationalise and do not facilitate an intuition about the underlying
structure of the network.  Various forms of consensus clustering
\citep{lancichinetti2012,goder2008} have been developed to optimise
non-overlapping modular decomposition ``on average'' over an ensemble
of datasets or runs of a non-deterministic community detection
algorithm.  However, it remains debatable whether these communities
represent the ``true'' communities of the network, or just the best
possible consensus solution given the algorithm and the available
data.

Our approach to the issue of community ambiguity is predicated on the
observation that, although the community structure of a network may
not be certainly known, there will generally be variability between
nodes in terms of the certainty with which they can be individually
affiliated with a specific community.  Here, we seek to formalise this
intuition by developing a new metric called \emph{versatility} which
can be used to quantify the certainty with which each node is assigned
to the same community of a network across multiple runs of the
community detection algorithm on the same network.

In what follows, we first define an estimator of nodal versatility of
community affiliation and demonstrate its desirable properties by
analysis of idealised networks designed for maximal community
ambiguity. To build intuitive understanding of what versatility is
measuring, we explored its performance in two real-life networks: the
karate club graph, a social network; and the mouse brain connectome, a
brain network derived from anatomical tract-tracing experiments. In
both of these cases, we show how versatility can be used to identify
the resolution parameters of the Louvain hierarchical community
detection algorithm \citep{blondel2008} that provide the least
ambiguous modular decomposition of the network as a whole.
Additionally, we use versatility characterise the topological roles of
each individual node.

\section{Methods}

We propose a measure of nodal versatility of community affiliation
that can be estimated for any graph (weighted or unweighted, directed
or undirected), and for any non-deterministic or stochastic algorithm
which decomposes the community structure of such graphs.

\subsection{Definition of versatility}

The most natural way to accurately capture the intuition of
variability in community classification with respect to an arbitrary
algorithm applied to a single graph is to run a stochastic community
detection algorithm many times. Because the partitions will be
different, we use as our fundamental quantity the probability
$p_{i,j}$ that any two nodes $i$ and $j$ will be classified in the
same community. This is equivalent to the element in the $i$'th row
and the $j$'th column of the association matrix from consensus
clustering \citep{lancichinetti2012}. If two nodes are always in the
same community, $p_{i,j}$ will be equal to 1, and if they are never in
the same community, it will be 0. Likewise, if they are in the same
community 50\% of the time, $p_{i,j}$ will be equal to 0.5.
Versatility should be highest for a node $j$ when $p_{i,j}=0.5$ for
all $i \neq j$.  To formalise this idea, we transform the $p_{i,j}$ values
with the sine function, and then sum the transformed $p_{i,j}$ values
for each node $j$.  Finally, we normalise by the average number of
nodes in the community containing node $j$.  So, for node $j$, we sum
$\text{sin}(\pi p_{{i, j}})$ for all nodes $i$ and normalise by
dividing by the mean size of the communities containing $j$ (weighted
by membership probability).

More formally, the versatility of a node $j$ is defined as

\begin{equation}
V(j) = \frac{\sum_{i} \sin(\pi \E(a(i, j)))}{\sum_{i} \E(a(i, j))} \label{eq:vers}
\end{equation}

\noindent where $\E{}$ is expected value, and 

\begin{equation*}
  a(i, j)  = \begin{cases} 1, & i \text{ and } j \text{ are in the same community} \\ 0, & \text{else} \end{cases}
\end{equation*}

\begin{figure}
\centering
\includegraphics[width=\textwidth]{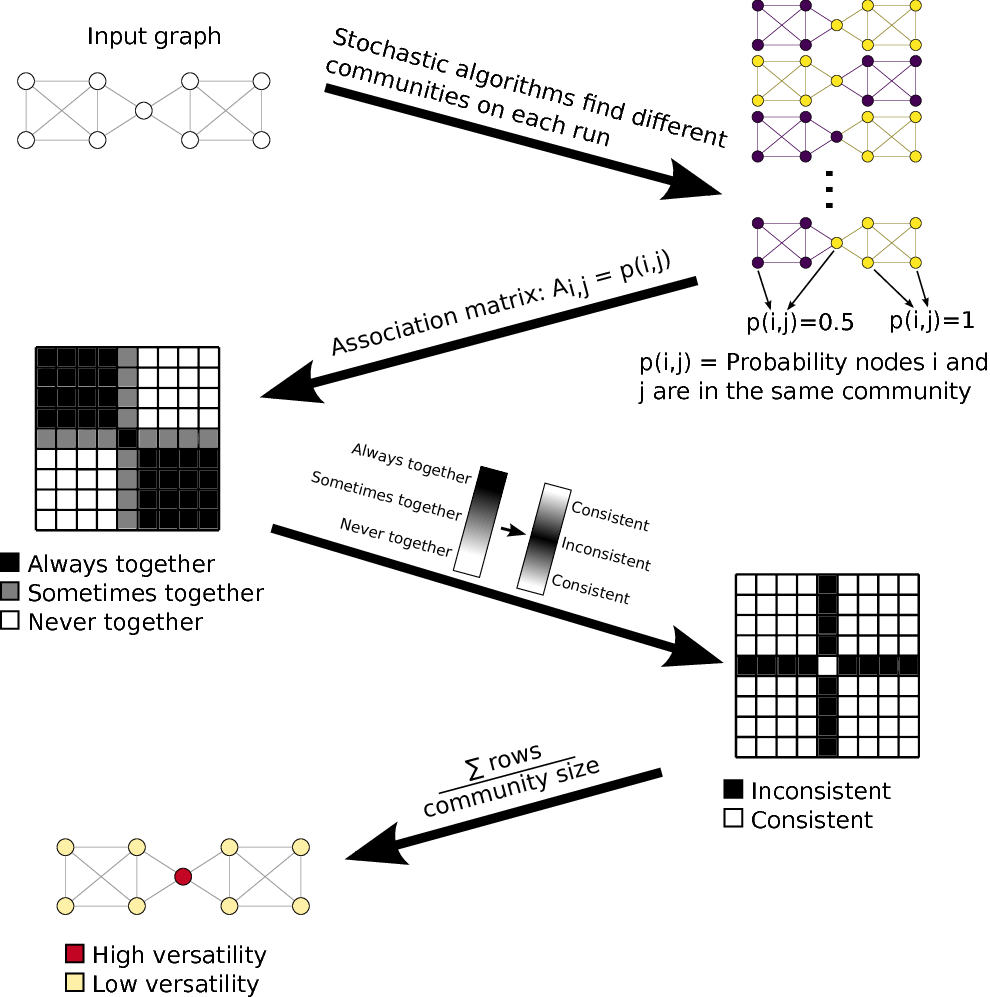}
\caption{\textbf{A schematic overview of versatility.}  A stochastic
  community detection algorithm is run many times, generating a
  collection of graph partitions.  Each pair of nodes corresponds to a
  cell in the association matrix, which describes the sample
  probability that any two nodes will be grouped in the same
  community.  A transformation function is applied to this matrix, so
  that pairs of nodes that are consistently grouped either in the same
  community or in a different community are given low values, and
  pairs which are only sometimes in the same community are given high
  values.  The sum is taken for each node's possible pairs, and
  normalised by the mean size of the communities weighted by the
  node's community membership.  The resulting normalised sum is the
  node's versatility. \label{fig:fig0}}
\end{figure}

\noindent Because $\E(a(i,j))=p_{i,j}$, we estimate the expected value
of $a(i, j)$ by running the community detection algorithm many
times. See Fig \ref{fig:fig0} for a graphical summary of this
process.

This formula for versatility can be computed for any stochastic
community detection algorithm. If versatility is being used to
understand a community decomposition, the same algorithm and
parameters should be used for each iteration as for the original
decomposition. For algorithms based on Newman's quality function $Q$
\citep{newman2004} such as the Louvain algorithm \citep{blondel2008},
or for any other algorithm with a resolution parameter, it is useful
to compute versatility for many different resolution parameters and
then take the average.  When multiple potential resolution parameters
could be chosen, this increases the stability of versatility removes
the effects that are specific to a single resolution parameter.

For the Louvain algorithm, we have determined that approximately 1000
runs at a particular resolution parameter gives reliable results;
however, in practice, for most algorithms including the Louvain
algorithm, 200 iterations seems to give a reasonable estimation based
on numerical simulations tracking the variance in versatility across
multiple runs in the same network (Fig S1).  Unless otherwise
specified, we compute versatility by taking the mean Louvain
versatility across a spectrum of resolution parameters from
$\gamma\in\{0.4,0.5,\ldots,2.4\}$ with 100 iterations each, for a total
of 2100 iterations.

Code to calculate versatility is available in Python and
\textsc{Matlab}/Octave from
\url{https://github.com/mwshinn/versatility}. This includes code to
compute the average versatility across a spectrum of resolution
parameters, as well as to generate a plot which can be used to find an
optimal resolution parameter.

\subsection{Technical evaluation of versatility estimators}

\begin{figure}
\centering
\includegraphics[width=\textwidth]{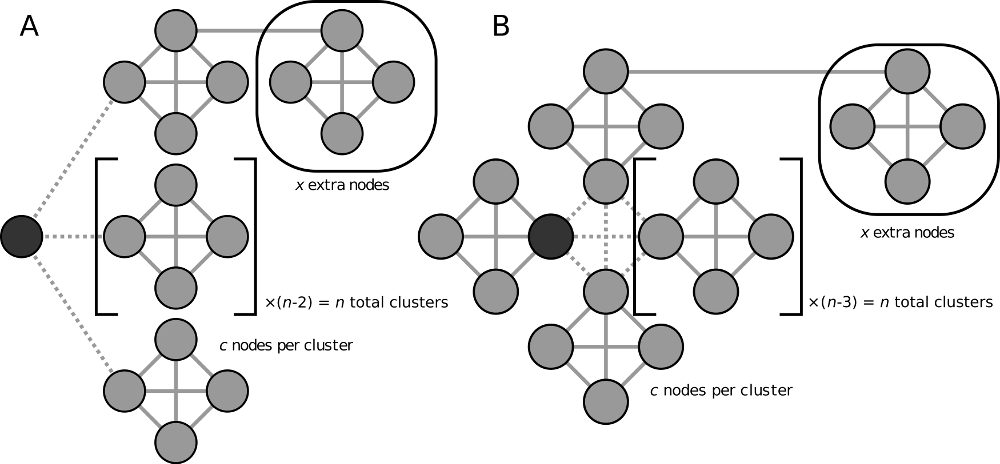}
\caption{\textbf{Test networks for evaluation of candidate versatility
    metrics.} (a) The n-split case. A indicator node (coloured darker)
  can be affiliated with exactly one of the $n$ clusters with
  probability $1 / n$. (b) The n-clusters case. There are $n$ clusters
  each with a probability $p$ of connecting to any other given
  cluster. The indicator node is a member of one of these clusters
  which is weakly connected to other clusters. In both cases clusters
  are of size $c$, and these cases are assumed to be a part of a
  larger network, with $x$ other nodes. \label{fig:W}}
\end{figure}

In theory, there are several ways in which versatility of nodal
affiliation to a modular community structure could be defined. To
choose between the many versatility estimators that are potentially
available, we first list the desirable properties of a theoretically
optimal estimator and then evaluate a number of candidate estimators
against these criteria, using two test networks to assess the
performance of each estimator empirically, as shown in Fig
\ref{fig:W}.

\subsubsection{Model networks}

The first model is called the ``n-split network'' and represents the
case where a single \emph{indicator node} has one connection to each
of a number of identical tightly-interconnected clusters. It has three
parameters: $n$, the number of clusters to which the indicator node is
connected; $c$, the number of nodes in each cluster; and $x$,
additional nodes in the network that change the total network size but
are assumed to not affect the community detection. Thus, the total
size of the network is $n \times c + x + 1$. We assume that the
indicator node can be assigned to only one community by each run of a
stochastic community detection algorithm, with the probability of affiliation
to each model on each run being equal to $1 / n$.

The second model is called the ``n-clusters network'' and represents
the case where there are several identical clusters in the network
which are each tightly interconnected within themselves, but only
loosely (and symmetrically) connected to each other.  This model has
four parameters which echo those of the n-split case: $n$, the number
of clusters we are to consider in the network; $c$, the number of
nodes in each cluster; $x$, the number of additional nodes in the
network that change the total network size but are assumed to not
affect the community detection; and $p$, the probability that, in any
given run of an ideal community detection algorithm, any two clusters
will be grouped in the same community.  The indicator node is taken to
be an arbitrary node in an arbitrary cluster, which are assumed to
have identical properties due to symmetry.  The indicator node is
connected to a cluster with a total of $c$ nodes, and this
\emph{indicator cluster} is connected with probability $p$ to $n-1$
other clusters each of size $c$. In any given run of an ideal
community detection algorithm, the indicator cluster can be assigned
its own unique community or it can be affiliated to a larger community
also comprising one or more of the other clusters, each with
probability $p$. Thus the indicator cluster will be affiliated to the
same community as at least one other cluster with probability
$1 - (1 - p)^{n}$, and it will be affiliated to the same community as
all of the other clusters with probability $p^{n}$.

It is important to note that these networks are strictly
theoretical. It is unknown whether any given community detection
algorithm would actually exhibit the partitions described above. For
example, in practice, the Louvain algorithm behaves deterministically
in the n-split case, preferring always to affiliate the indicator node
with the same community based on the algorithmic bias caused by the
order of the adjacency matrix.  However, the behaviours described
above are what we would expect if the model networks were decomposed
into a community structure by a stochastic community detection
algorithm which exhibits the idealised behaviours described in
conjunction with each network; for example, in practice, most
algorithms show biases based on the ordering of the adjacency matrix,
and hence may not exhibit such behaviour.

\subsubsection{Desirable properties of versatility metrics}

In this context, we can list the key desirable properties of a
versatility metric as follows:

\begin{description}
\itemsep1pt\parskip0pt\parsep0pt
\item[Lower bound] Versatility should be bounded below by 0.
\item[Upper bound as network size $\rightarrow \infty$] A universal upper bound
  on versatility should exist regardless of network size.
\item[Degree invariance] The degree of a node, or the number of
  connections it makes, should not directly affect its versatility.
\item[n-split zero] If the number of clusters $n = 1$ in the n-split
  network, the versatility of the indicator node is 0.
\item[n-cluster zero] If the probability of being grouped with another
  cluster $p = 0$ or $1$ in the n-cluster network, the versatility of
  the indicator node is 0.
\item[Network size invariance] Versatility is unaffected by the total
  number of nodes in the network, represented by $x$ in both the
  n-split and n-cluster networks.
\item[n-split cluster size monotonicity] Versatility is an increasing
  function of the number of nodes per cluster $c$ in the n-split
  network.
\item[n-cluster cluster size invariance] Versatility is unaffected by
  the number of nodes per cluster $c$ in the n-cluster network.
\item[Cluster number monotonicity] Versatility is an increasing
  function of the number of clusters $n$ in both networks.
\item[Splitting over breaking] All else being equal, a node
  probabilistically connected to two communities should have higher
  versatility than a node in a cluster that is probabilistically
  contained within the same community as other clusters. So for
  otherwise equal networks, the n-split case should result in higher
  versatility than the n-cluster case.
\end{description}

\subsubsection{Candidate versatility metrics}\label{potential-metrics}

We evaluated six candidate versatility metrics that were expected to
be reasonably well-behaved. They followed the general form

\begin{equation}
V(j) = \frac{1}{g(j)} \sum_{i} f(\E(a(i, j)))
\end{equation}
\noindent where $a(i, j)$ is a stochastic function that measures whether nodes
$i$ and $j$ are in the same community; $g : [0, 1] \rightarrow (0,
\infty)$ is a normalisation function; $\E$ is the expected value; and
$f : [0, 1] \rightarrow [0, 1]$ is any continuous, concave function
that has the values $f(0) = 0$, $f(1) = 0$, and $f(0.5) = 1$, and is
symmetric around $0.5$, i.e. $f(x) = f(1 - x)$.

Three functions were chosen for $f$ and two for $g$, and each was
denoted by a capital letter; the six possible combinations of $f$ and
$g$, each denoted by a two-letter code, represented the six
versatility metrics tested. For $f$, we evaluated the sine function,

\begin{equation*}
f(x) = \text{sin}(\pi x) \equiv \texttt{S}
\end{equation*}

\noindent the entropy function,

\begin{equation*}
f(x) = - x \log_{2}(x) - (1 - x) \log_{2}(1 - x) \equiv \texttt{E}
\end{equation*}

\noindent and the triangle wave function,

\begin{equation*}
f(x) = \begin{cases} x, & 0 \leq x \leq .5 \\ 1 - x, & .5 \leq x \leq 1 \end{cases} \equiv \texttt{T}
\end{equation*}

For $g$, we tried normalising by the number of nodes in the network

\begin{equation*}
g(j) = \sum_{i} 1 \equiv \texttt{U}
\end{equation*}

\noindent and also by the mean community size

\begin{equation*}
g(j) = \sum_{i} \E(a(i, j)) \equiv \texttt{C}
\end{equation*}

While we could not examine the full space of potential functions $f$
and $g$, we considered those that seemed most natural given the
constraints.  We did not consider metrics that were normalised by the
degree of the node, because this breaks the intuition that versatility
should depend only on the community classification. We also did not
explore metrics which depend only on the probability of the indicator
node being in the same community as its nearest neighbours, rather than
all nodes, as these by definition do not satisfy the degree invariance
property.  Such metrics may be more suitable as a community-agnostic
version of participation coefficient, rather than assessing the degree
to which a node affiliates with a community.

\subsubsection{Participation coefficient}

Versatility was contrasted with participation coefficient
\citep{guimera2005}, because in informal terms, both describe the
coupling of a node with its community.  Participation coefficient
tries to measure the intuitive property of whether nodes could
facilitate communication across separate groups of nodes by having a
high inter-community degree.  It is defined as

\begin{equation}
  \mbox{PC}(i) = 1 - \sum_{s=1}^K\left(\frac{k_{i,s}}{k_i}\right) \label{eq:pc}
\end{equation}

\noindent where $K$ is the number of communities, $k_i$ is the degree
of node $i$, and $k_{i,s}$ is the intra-community degree of node $i$.

\subsubsection{Performance of candidate metrics}

Each of the 6 candidate versatility metrics was benchmarked by its
performance in analysis of the two model networks as shown in Table
\ref{tab:ST1}, where they are compared to participation coefficient
(PC).

\begin{table}
\centering
\caption{\textbf{A list of desirable properties satisfied by each of
    the algorithms.} ``\CM'' indicates satisfying the property. Numeric
  values are listed where relevant. * denotes trivially satisfying the
  monotonicity properties by being constant functions. {\dag} denotes
  that participation coefficient trivially satisfies certain
  properties due to the lack of degree invariance. Upper bounds were
  found analytically.\label{tab:ST1}}
\begin{tabular}{@{}lllllllll@{}}
Metric & \SU & \EU & \TU & \SC & \EC & \TC & PC  \\\hline
Lower bound & 0 & 0 & 0 & 0 & 0 & 0 & 0   \\
Upper bound as network size $\rightarrow \infty$ & 1 & 1 & 1 & $\pi$ & $\infty$ & 2 & 1  \\
Degree invariance & \CM & \CM & \CM & \CM & \CM & \CM &   \\
n-split zero & \CM & \CM & \CM & \CM & \CM & \CM & \CM  \\
n-cluster zero & \CM & \CM & \CM & \CM & \CM & \CM &  \\
Network size invariance & & & & \CM & \CM & \CM & \CM  \\
n-split cluster size monotonicity & \CM & \CM & \CM & \CM & \CM & \CM & *  \\
n-cluster cluster size invariance & \CM & \CM & \CM & \CM & \CM & \CM &  \\
n-split cluster number monotonicity & & & & \CM & \CM & * & \dag  \\
n-cluster cluster number monotonicity & \CM & \CM & \CM & \CM & \CM & \CM & \dag  \\
Splitting over breaking & \CM & \CM & \CM & \CM & \CM & \CM &   \\
\end{tabular}
\end{table}

Normalisation by the size of the network (\texttt{U}) was not
effective. None of the metrics \SU, \EU and \TU satisfied the
desirable property of network size invariance. Thus these metrics were
not considered further. The three metrics normalised by community size
(\SC, \EC, \TC) were much more evenly matched. Because \TC only
trivially satisfies n-split cluster monotonicity, we consider only \SC
and \EC. In practice, \SC and \EC give nearly identical results in all
networks for which they have been computed. \SC and \EC are very
similar and would both make good measures of versatility; however, we
select \SC for two reasons. First, and most importantly, it has upper
bound. Second, it avoids the potential for confusion created by using
the formula for Shannon's entropy, when in fact no
information-theoretic quantity is involved \citep{shannon1956}.  Thus,
\SC is equivalent to Equation \ref{eq:vers}.

While the versatility of any given node depends only on community
affiliation and does not depend on the size of the network, one
intriguing aspect of \SC is that the maximum possible versatility a
node is capable of achieving in a network does depend on the network
size.  This is due to the fact that increasing the size of the network
increases the number of nodes that could potentially be in the same
community as the indicator node.  In finite networks with $N$ nodes,
we can construct a network which maximises versatility of the
indicator node by considering a n-split network with clusters of size
$c=1$, with a total of $n=N-1$ clusters and $x=0$ extra nodes.  Fig S2
shows the maximum versatility as a function of network size.  Only for
networks with infinitely many nodes can \SC reach its maximum of
$\pi$.

We elected not to normalise versatility by $\pi$ because the maximum
versatility in any network with a finite number of nodes will be less
than $\pi$, with the limit depending on the number of nodes.  Thus, it
would be misleading to imply that the maximum versatility is 1 in a
finite-sized network.  Furthermore, versatility in the un-normalised
form has other distinctive landmarks.  For example, a versatility of
2.0 means that a node is perfectly split between two equally-sized
communities (as the community size approaches infinity).

We also chose not to normalise versatility by the upper bound given
the network size (the curve in Fig S2).  Doing so would create a
dependence on network size, and thus no longer satisfy the desirable
properties listed above.  An additional problem with normalising is
that the maximum versatility given the network size currently must be
calculated numerically, and thus the difficulty of implementing
versatility would increase.  Furthermore, a conceptual problem with
normalising by the maximum given the network size is that it cannot
possibly be as difficult to classify a node in a 2-node network as it
is to classify a node in a 1000-node network; in the former case there
are only two possible partitions for the network, whereas in the
latter there are approximately $10^{1927}$ (the 1000th Bell
number).

Thus, due to the desirable properties of \SC, the definition of
versatility is as given in Equation \ref{eq:vers}.

\section{Results}

\subsection{Versatility can find optimal modular resolution parameters}

Initially, Girvan and Newman's $Q$ modularity did not include a
resolution parameter \citep{newman2004}; only later was such a
parameter $\gamma$ added to search for community structure across a
range of community sizes, from a few large communities (low $\gamma$)
to a larger number of smaller communities (high $\gamma$)
\cite{reichardt2006}.  As a result, $Q$ values are highly dependent on
the resolution parameter.  The scaling of $Q$ also depends on the
value of $\gamma$---the modular resolution parameter---at which it is
calculated.  Thus, due to this bias, it is not possible to maximise
$Q$ across values of $\gamma$ and obtain a result relevant to the
structure of the network, making it difficult to use $Q$ to define an
``optimal'' resolution of the hierarchical community structure or
value of $\gamma$.

By contrast, a network's mean versatility depends only on how
consistently the nodes in the network affiliate with a specific
community.  Thus, we could define the ``optimal'' value of $\gamma$ as
that for which the average versatility is lowest, or for which the
community structure of the network is least ambiguously defined.
Newman \citep{newman2016} developed a theoretical formulation of an
optimal resolution parameter, but the proposed iterative procedure is
only guaranteed to converge when the network communities are drawn
from a special case of the stochastic block model \citep{condon2001}.
The proposed versatility-based method works on all networks, without
assumptions about the form of their community structure, and also on
non-$Q$ metrics that vary as a function of a resolution parameter.

When the global mean versatility is plotted at each value of $\gamma$
within a reasonable range, it varies as a function of the resolution
parameter and there are typically one or more values of $\gamma$
corresponding to local minima in global mean versatility.  At the
extremes of the curve, the versatility will be zero, as these
represent the cases of either a single community encompassing the entire
network, or each node in the network being in a separate community.
While these parameter ranges are the global minimum of versatility,
they are not desirable because they do not take into account the
purpose of the community decomposition itself, i.e. to find useful
communities.

In order to balance practical considerations with a principled method
of using the versatility curve to select optimal $\gamma$, several
approaches may be taken.  One can choose the value of $\gamma$ which
globally minimises the ambiguity in modular classification among those
values of $\gamma$ which generate non-trivial community structure.  In
other words, this finds a community structure which maximises the
ability of the algorithm to assign nodes to communities.  One can also
choose the value of $\gamma$ within a range of resolution parameters
that gives a theoretically expected number of communities, or which
satisfy another practical requirement.  For instance, in brain
networks where multiple modalities can be used to define different
networks in the same subject \citep{bullmore2009}, it may be most
useful to find equal numbers of communities in each subject to compare
across modalities.  Additionally, many networks exhibit a region of
$\gamma$ values for which versatility is consistently low across many
nearby resolution parameters.  Even if it does not globally minimise
versatility, this scheme prioritises the stability of the mean
versatility across small perturbations in resolution parameter.

The important point is that the global mean versatility curve provides
an objective function to guide the otherwise unrestricted and
unprincipled choice of resolution parameters often corresponding to
different community structures.  By understanding which resolution
parameters minimise ambiguity, we can make a more informed and precise
selection.

\subsection{Evaluation on the karate club network}

Zachary's karate club graph \citep{zachary1977} is a non-trivial
benchmark and standard test case for community detection
algorithms. This network represents friendships in a university karate
club before a political conflict caused the club to split into two:
members of the club are nodes and friendships between members are
edges.  Most community detection algorithms are able to find two
distinct groups of individuals in the club which correspond to the two
political factions in the club after the split.  However, there is one
individual who has exactly one friendship on one side of the split and
one friendship on the other side.  Community detection algorithms are
forced to assign this individual node to only one of the two
communities.

\begin{figure}
\centering
\includegraphics[width=\textwidth]{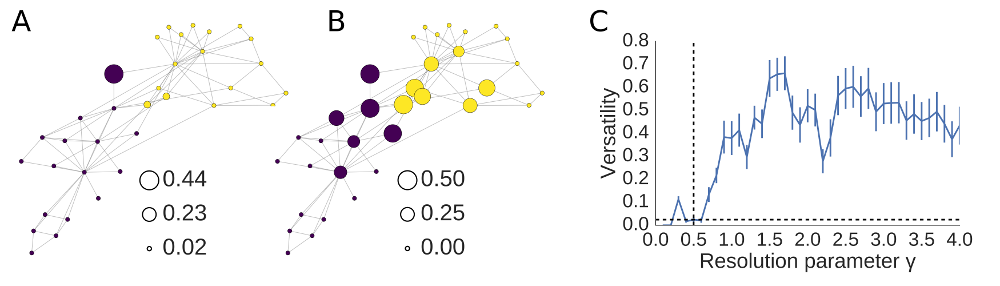}
\caption{\textbf{Versatility behaves as expected in the karate club
    network.} (a) The versatility in Zachary's karate club network is
  compared to (b) the participation coefficient in the same network,
  where the size of the node represents the versatility or the
  participation coefficient, respectively.  Versatility is only high
  for the nodes in between communities, whereas participation
  coefficient is also high for the hubs since they tend to have more
  edges into the other community.  The node with very high versatility
  has exactly one edge in each community.  Nodes are coloured
  according to their community with a resolution parameter of 0.5,
  which is shown in (c) to minimise the mean versatility,
  i.e. providing the most stable communities.  Participation
  coefficient in (b) was also calculated according to this partition
  structure.  We know \textit{a priori} that this club split from one
  group into two factions, and indeed this range of resolution
  parameter gives two communities.  Error bars represent
  SEM.
  \label{fig:karate}}
\end{figure}

As we see in Fig \ref{fig:karate}a, most of the nodes have low
versatility, except the individual with one friendship in each of the
communities.  Nodes that are connected to only one group cleanly sort
into their respective faction.  Versatility distinguishes itself from
participation coefficient shown in Fig \ref{fig:karate}b by only
holding a high value for the nodes which could be classified into
either community by the algorithm.  Participation coefficient, by
contrast, highlights the degree to which a node is an inter-community
hub.  We might expect the individuals with high participation
coefficients to make effective mediators in this conflict or to be
ambassadors between the factions.  By contrast, we would expect the
individual with high versatility to have a difficult decision on which
of these two clubs to join after the split occurred.  This exemplifies
an important difference between participation coefficient and
versatility.

When we look at the curve of versatility across different resolution
parameters $\gamma$ in Fig \ref{fig:karate}c, we see versatility is
near-zero around $\gamma=0.5$.  This happens to be a parameter which
divides the network into two communities, consistent with the prior
knowledge that this social network was indeed divided into two
communities \citep{newman2006}.

\subsection{Mouse connectome}

With recent advances in biotechnology, it has become possible to use
graph theory based techniques to study the brain \citep{bullmore2009}.
Such work has shown promise in helping us understand concepts ranging
from locomotion in model organisms \citep{towlson2013} to diseases as
complex as schizophrenia \citep{lynall2010}.  Since the brain has long
been hypothesised to function as a set of semi-independent modules
\citep{fodor1983}, and the communities of brain networks obtained
using graph theory have anatomical and functional significance
\citep{chen2008}, it is only natural to talk about brain modules as
communities in brain networks \citep{sporns2016}.

While there are many ways to find which regions of the mammalian brain
are connected, one of the most reliable ways is by injecting a
fluorescent viral tracer into a source region of the brain.  In the
days following injection an anterograde viral tracer travels along the
axons of neurons projecting from the source region to anatomically
connected target regions.  By measuring the strength of the
fluorescent tracer in high resolution microscopic images of the
injected animal's brain, it is possible to quantify the weight of
anatomical connectivity from a source region to each possible target
region.  By performing this experiment in many different mice, with
different source regions injected in different experiments, it is
possible to estimate the complete anatomical connectivity matrix or
connectome of the mouse brain \citep{oh2014}.

A weighted, directed network was constructed from 112 brain regions of
the mouse brain connectome derived from over 400 of such tract-tracing
experiments conducted by the Allen Institute for Brain Sciences \citep{oh2014}, as
previously described \citep{rubinov2015}.

\begin{figure}
\centering
\includegraphics[width=\textwidth]{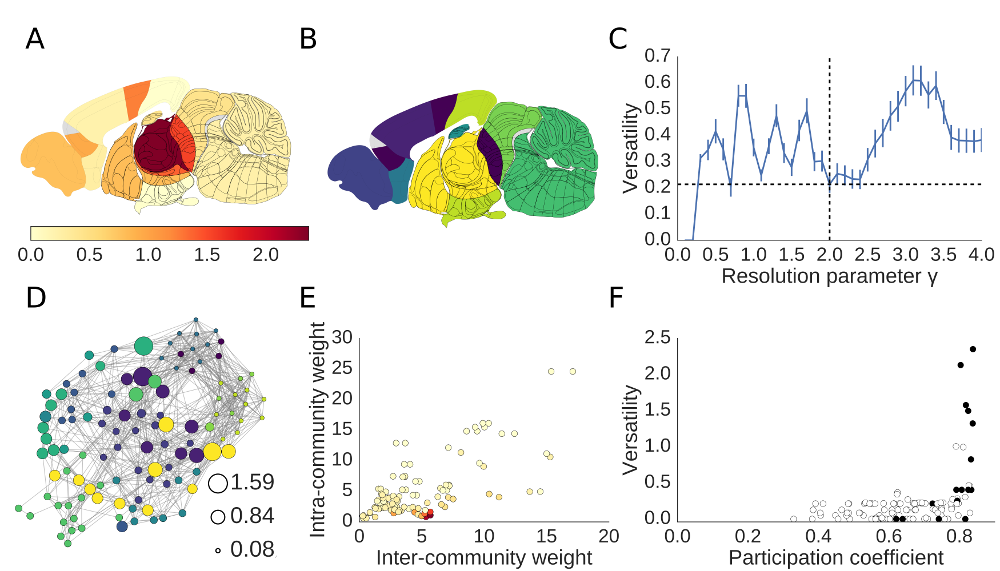}
\caption{\textbf{Versatility in the mouse connectome.} (a) The
  versatility of each region in the mouse brain, plotted anatomically.
  (b) An example classification of the communities in mouse at an
  optimal resolution parameter, determined using the curve shown in
  (c).  Error bars represent SEM.  (d) A topological view of the mouse
  connectome.  Nodes are coloured by community using the colour scheme
  from (b).  The mean versatility is given by the size of the
  node. (e) Versatility is related to inter-community weight and
  intra-community weight.  Highly versatile nodes have low
  intra-community weight and high inter-community weight.
  Versatility, indicated by colour, is the mean versatility across
  resolution parameters.  Colours correspond to the colour bar in
  (a). (f) Versatility is plotted against participation coefficient,
  where participation was computed at the resolution parameter from
  (c).  Previous work \citep{rubinov2015} identified several nodes,
  denoted ``hi-par'' nodes (meaning ``high participation''), which are
  coloured black.  The mean versatility of the hi-par nodes is
  significantly (Wilcoxon-Mann-Whitney $p < .001$) different from the
  non--hi-par nodes. \label{fig:fig2}}
\end{figure}

The communities of this network were found after using versatility to
choose an optimised resolution parameter ($\gamma = 2.0$) and are
displayed in anatomical coordinates in Fig \ref{fig:fig2}b. At this
resolution parameter, there are 11 modules across the two hemispheres
with similar specialisations to those found previously
\citep{rubinov2015}.  Rather than select the resolution parameter
corresponding to the global minimum of versatility across non-trivial
partitions, we selected a value which had consistently low versatility
across local perturbations in resolution parameter.  The anatomical
map of versatility (averaged across resolution parameters) is shown in
Fig \ref{fig:fig2}a.  Fig \ref{fig:fig2}d shows the topology of the
network, and demonstrates that there are versatile nodes in all
communities; versatility is not concentrated in a few communities. the
versatile nodes do not cluster around any individual communities.

This network was previously found \citep{rubinov2015} to have a
hierarchical community structure, such that a few large functionally
specialised communities were subdivided into smaller sub-communities
as the resolution parameter of the Louvain community detection
algorithm was incrementally increased.  This means that small,
fine-scale modules were associated with larger, coarse-grained modules.
For example, at fine scales (with higher resolution parameters), the
auditory and visual modules were separate, but at higher scales (with a
lower resolution parameter) they combined to form the audio-visual
module.  Most nodes were consistently affiliated to the same
community, or one of its offspring sub-communities, over the spectrum
of resolution parameters. However, a subset of nodes which also tended
to have high participation coefficient (giving them their namesake
\emph{hi-par} nodes) were inconsistently affiliated to
(sub-)communities in this hierarchy.

Due to the similarity between the criteria for hi-par nodes and the
definition of versatility, we hypothesised that the hi-par nodes would
have a higher versatility than the non-hi par nodes.  As shown in
Fig \ref{fig:fig2}f, versatility is significantly higher for hi-par
nodes than it is for non--hi-par nodes; in other words, nodes with
high versatility do not fit well into the community hierarchy.  This
includes nodes in the diencephalon, prefrontal cortex, and basal
ganglia; see Fig S3 and Table S1 for details.

\section{Discussion}

In a world where community structure is not as black and white as it
promises to be, more information than a discrete partition is
needed. Community detection algorithms struggle to find a set of
non-overlapping communities that adequately represent the community
affiliations of the nodes in the network. Furthermore, it is not even
clear epistemologically that there \emph{is} a reasonable underlying
community structure to any given network; at the extreme end, there is
certainly no community structure to an Erd\H{o}s--R\'{e}nyi random
network, though many algorithms will still yield communities. This is
problematic when some of the analyses performed on networks depend
highly on the community affiliation of a node, such as the dynamic
community structure metrics \citep{bassett2011,mattar2015}.  This
complicates analysis, especially when the change in a node's community
affiliation is the variable of interest.

Rather than seeing this stochasticity as a disadvantage, we can use it
to extract previously discarded information about which nodes are not
closely affiliated with any communities.  To this end, we developed a
measure called versatility which satisfies a set of desirable
theoretical properties.  We explored how versatility can be used to
choose a resolution parameter for a community detection algorithm based
on how much each parameter reduces the ambiguity of the community
structure.  Finally, we examined the versatility of nodes in two
networks: the karate club network, and the mouse brain connectome.
The karate club network and the mouse brain connectome were both
consistent with previous work describing the nodes that do not fit
well into any of the modular communities.

Versatility has two core applications for which it is useful.  First
when used in conjunction with a particular community structure, it is
useful as a post-hoc method to capture the information about the
reliability of community assignment.  In this sense, versatility can
be seen as a description of the interaction between the algorithm and
the network.  When using versatility in this way, it is generally
desirable to match the algorithm (and the resolution parameter, if
applicable) to the network being analysed so that it is possible to
determine the reliability with which the information on partition
assignment can be interpreted.  Second, it is useful in its own right,
as a way of finding nodes that do not fit very well with any
community.  In many cases, these nodes are the most interesting
because their interactions are not representative of the community to
which they belong, and no community is a good predictor of their
interactions.  The interpretation of versatility in this case is
highly dependent on the particular network.

Unlike most community-based methods, versatility is not limited to
communities defined by strong intra-modular connections and weak
inter-modular connections.  For example, previous authors
\citep{arenas2008,benson2016} have defined communities based on the
similarity between motif profiles.  Under these definitions, it is
possible for a community to have \textit{no} intra-community edges.
In abstract community definitions, versatility works identically
without modification.

Participation coefficient and versatility share superficial
similarities but are conceptually very different.  Participation
coefficient is highly dependent on the particular partition specified
by the community detection algorithm, and can often give unintuitive
results.  This is because participation coefficient relies on the
assumptions that ``true'' underlying communities exist, and that they
can be detected by the algorithm.  If either of these conditions
doesn't hold, the participation coefficient is difficult to interpret.
Participation coefficient was designed for a different
purpose---detecting which nodes are important for inter-community
communication and linkage---and is able to find inter-community hubs
and distinguish them from intra-community hubs \citep{guimera2005}.
However, it is not successful in uncovering the certainty with which
each node can be assigned to a community.

In the words of Kuhn \citep{kuhn1964}, ``The decision to employ a
particular piece of apparatus and to use it in a particular way
carries an assumption that only certain sorts of circumstances will
arise.''  Because we have access to algorithms that will separate a
network into communities, it is easy to assume that the communities
found using these algorithms are ``the true underlying communities''
and that these communities should be taken as truth in the further
analysis of the network.  But of course, as we have seen, some
assignments may have more truth to them than others.

\section{Acknowledgements}

M.S.\ was supported by the Winston Churchill Foundation of the United
States.  J.S.\ was supported by the NIH Oxford-Cambridge Scholars
Program.  F.V.\ was supported by the Gates Cambridge Trust.  Thank you
to Rolf J.F.\ Ypma for for assistance with the mouse connectome.
Computing support was provided by the NIHR Cambridge Biomedical
Research Centre.

\section{Author contributions statement}

All authors conceived the experiments, M.S.\ conducted the experiments
and analysed the results, M.S.\ and E.B.\ wrote the paper, all authors
reviewed the manuscript.

\section{Financial disclosure}

E.B.\ is employed half-time by the University of Cambridge and
half-time by GlaxoSmithKline (GSK); he holds stock in GSK.

\end{document}